# Unrevealing hardening and strengthening mechanisms in high-entropy ceramics from lattice distortion

Yiwen Liu[1], Haifeng Tang[1], Mengdong Ma[1], Hulei Yu, Zhongyu Tang, Yanhui Chu*

School of Materials Science and Engineering, South China University of Technology,

Guangzhou, 510641, China

[1] These authors contribute equally to this article

* Corresponding author. chuyh@scut.edu.cn (Y. Chu)

**Abstract**

Revealing the hardening and strengthening mechanisms is crucial for facilitating the design of superhard and high-strength high-entropy ceramics (HECs). Here, we take high-entropy diborides ($HEB_2$) as the prototype to thoroughly investigate the hardening and strengthening mechanisms of HECs. Specifically, the equiatomic 4- to 9-cation single-phase $HEB_2$ ceramics (4-9$HEB_2$) are fabricated by an ultra-fast high-temperature sintering method. The as-fabricated 4-9$HEB_2$ samples possess similar grain sizes, comparable relative densities (up to ~98%), uniform compositions, and clean grain boundaries without any impurities. The experimental results show that the hardness and flexural strength of the as-fabricated 4-9$HEB_2$ samples have an increasing tendency with the increase of metal components. The first-principles calculations find that lattice distortion is essential to the hardness and strength of $HEB_2$. With the increase of metal components, an aggravation of lattice distortion accompanied by B-B bond strengthening is determined, resulting in the enhancement of the hardness and flexural strength. Moreover, the correlation between other potential indicators and the hardness/flexural strength of $HEB_2$ has been disproved, including valence electron concentration, electronegativity mismatch, and metallic states. Our results unravel the hardening and strengthening mechanisms of HECs by intensifying lattice distortion, which may provide guidance for developing superhard and high-strength HECs.

## 1. Introduction

High-entropy ceramics (HECs) are often defined as multicomponent inorganic compounds with one or more Wyckoff sites shared by four or more principal elements and each element between 5 and 35 at.% [1,2]. Since their discovery in 2015, HECs have attracted intense interest in the ceramic community due to their vast compositional space, unique microstructural features, and tunable properties [3,4]. To date, many efforts have been devoted to developing different kinds of HECs, including high-entropy oxides [4-6], high-entropy diborides ($HEB_2$) [7-11], high-entropy carbides [12-16], and high-entropy nitrides [17-19]. Owing to the four core effects of high-entropy materials, including high-entropy, lattice distortion, sluggish diffusion, and cocktail effect, HECs have been shown to possess exceptional properties, such as good thermal stability [3,6,20], low thermal conductivity [1,12,14,21], improved mechanical properties [1,7,12,14,22,23], good corrosion and radiation resistance [11,24-26], and superior electrochemical characteristics [5,27,28]. Stimulated by these remarkable properties, HECs are becoming highly anticipated candidates for applications in functional and structural fields, especially in extreme environments.

The mechanical properties, as one of the performance of HECs, are crucial to their use as structural materials, especially in extreme conditions. Up to now, HECs have been developed to demonstrate some outstanding mechanical properties such as high or super hardness and modulus [1,7,17-19,29,30], excellent flexural strength [31-33], and outstanding high-temperature creep resistance [34]. For example, most researchers have found that the hardness and modulus of HECs, including $HEB_2$, high-entropy carbides, and high-entropy nitrides, are significantly higher than those of their binary components [1,7,18], mainly due to the aggravation of lattice distortion [17,18], the improvement of bond strength [19,29], or the decrease of valence electron concentration (VEC) [30]. Meanwhile, Feng *et al.* [31] reported the enhanced room- and high-temperature flexural strength of $(Hf_{0.2}Nb_{0.2}Ta_{0.2}Ti_{0.2}Zr_{0.2})B_2$ $HEB_2$ compared to their binary diboride components, which mainly resulted from the lattice distortion and solid solution strengthening effects. Similar phenomena have also been found in $(Hf_{0.2}Zr_{0.2}Ti_{0.2}Ta_{0.2}Nb_{0.2})C$ high-entropy carbides [32,33]. Moreover, Han *et al.* [34] observed a better high-temperature creep resistance in $(Ta_{0.25}Hf_{0.25}Zr_{0.25}Nb)C$ high-entropy carbides compared to their binary carbide components, which was primarily ascribed to the enhancement of the lattice distortion and the high-temperature thermodynamic stability. These outstanding mechanical properties of HECs should be mainly attributed to their solution hardening and strengthening mechanisms, which may derive from the changes in lattice distortion, bond strength, or VEC. In addition to lattice distortion, bond strength, and VEC, some other underlying hardening and strengthening mechanisms that have been found in high-entropy alloy and low-entropy ceramics may also have a significant influence on the mechanical properties of HECs, like the electronegativity difference ($\delta Xr$) [35] and metallic states [36]. Therefore, the research on the hardening and strengthening mechanisms of HECs is still in its infancy, and the essence of the hardening and strengthening mechanisms of HECs has not been well clarified so far, which largely hinders the design of the superhard and high-strength HECs.

$HEB_2$, as the member of HECs, is considered promising structural materials for extreme circumstances due to their ultrahigh melting point, tailorable thermo-mechanical performance, superior chemical inertness, and good high-temperature stability [7-11,20,28,29,31]. Herein, we choose $HEB_2$ as the representative to systematically investigate the hardening and strengthening mechanisms of HECs. To be specific, the equiatomic 4- to 9-cation single-phase $HEB_2$ ceramics (4-9$HEB_2$) of IVB, VB, and VIB groups are fabricated by an ultra-fast high-temperature sintering (UHTS) method. The reason for choosing the UHTS method is due to the fact that its ultrafast heating rates and ultrahigh temperatures facilitate the full sintering of the desired samples, minimizing the uncontrolled grain coarsening. In the as-fabricated 4-9$HEB_2$ samples, uniform compositions and clean grain boundaries without any impurities are gained, as well as comparable grain sizes and relative densities. As a result, the impact of these factors on the variation of hardness and flexural strength can be ignored in this study. It should be noted that the hardness and flexural strength of $HEB_2$ are increased with the increase of metal components. In order to reveal the hardening and strengthening mechanisms of $HEB_2$, the influence of five underlying factors is investigated by first-principles calculations, including lattice distortion, bond strength, VEC, $\delta Xr$, and metallic states. Lattice distortion is found to play a dominant role in the improvement of the mechanical properties in $HEB_2$ by strengthening B-B bonds. This study may provide guidance for developing the superhard and high-strength HECs.

## 2. Experimental methods

As listed in Table S1, 4-9$HEB_2$ samples were fabricated via UHTS method, namely $(Ta_{1/4}Nb_{1/4}Ti_{1/4}Zr_{1/4})B_2$ (4$HEB_2$), $(Ta_{1/5}Nb_{1/5}Ti_{1/5}Zr_{1/5}Hf_{1/5})B_2$ (5$HEB_2$), $(Ta_{1/6}Nb_{1/6}Ti_{1/6}Zr_{1/6}Hf_{1/6}Cr_{1/6})B_2$ (6$HEB_2$), $(Ta_{1/7}Nb_{1/7}Ti_{1/7}Zr_{1/7}Hf_{1/7}Cr_{1/7}Mo_{1/7})B_2$ (7$HEB_2$), $(Ta_{1/8}Nb_{1/8}Ti_{1/8}Zr_{1/8}Hf_{1/8}Cr_{1/8}Mo_{1/8}V_{1/8})B_2$ (8$HEB_2$), and $(Ta_{1/9}Nb_{1/9}Ti_{1/9}Zr_{1/9}Hf_{1/9}Cr_{1/9}Mo_{1/9}V_{1/9}W_{1/9})B_2$ (9$HEB_2$). The detailed fabrication processes of 4-9$HEB_2$ samples were described as follows: Commercially available metal diboride powders as the raw materials (Table S2-S3) were first weighed according to the equal molar ratio of metal elements and then mixed in an agate mortar. The mixtures were milled manually using an agate pestle for 30 min inside a glove box filled with argon protective atmosphere to prevent their oxidation. Afterward, they were compacted into cylindrical pellets with the dimensions of $\phi$16 mm × 3 mm under a uniaxial pressure of either 10 MPa for 3 min (4-6$HEB_2$) or 12 MPa for 2 min (7-9$HEB_2$). The green pellets were placed into a round graphite mould, and then inserted into a graphite strip. Subsequently, the device was vacuumed to below $5 \times 10^{-3}$ Pa and slowly introduced an argon protective atmosphere to atmospheric pressure. After that, alternating current (AC, 0-250 A) was applied to both ends of the graphite strip to swiftly generate a large amount of Joule heat to heat samples. The surface temperature of the graphite strip was measured by an infrared thermometer (SensorTherm GmbH M313, Germany) during the fabrication process.

To evaluate the effect of the grain size on mechanical properties, the well-polished surfaces of the as-fabricated samples were etched for 60 s by using an acid mixture (HF:

$HNO_3$: $H_2O$ = 1:1:3), and at least 50 grains were counted to ensure the validity of the average grain size results. Furthermore, the relative densities of the as-fabricated samples were calculated based on experimental and theoretical densities of the samples measured by Archimedes' method and the lattice constants from Rietveld refinement of the crystal structure, respectively. The Vickers hardness of the as-fabricated samples was tested by employing a Vickers indenter (HVS-30Z, Shanghai SCTMC Co. Ltd.) under different applied loads (0.1 kg, 0.2 kg, 0.3 kg, 0.7 kg, and 1 kg). Meanwhile, the nanoindentation tests were carried out by using a Nano-Indenter TM XP (MTS system Corp.) system with a Berkovich diamond indenter to measure the nanohardness and elastic modulus of the as-fabricated samples. The displacement of the indenter was set at 800 nm to eliminate the size effect. The nanohardness and elastic modulus were calculated by the Oliver and Pharr method [37]. Over 50 measurements were conducted to ensure statistical validity and minimize the microstructural and grain boundary effects. The flexural strength measurements of the samples (1.8 mm × 2.8 mm × 8 mm, with a span of 7 mm) were conducted using a universal tester (UTM2103, Suns Co., Ltd., China) equipped with a 1000 N load cell. The loading rate was 0.05 mm $min^{-1}$. At least five samples were tested for each $HEB_2$.

The phase composition of the as-fabricated samples was analyzed by X-ray diffraction (XRD, X′pert PRO; PANalytical, Netherlands). The refinement was performed with a general structure analysis system (GSAS) software. The microstructure and compositional uniformity of the as-fabricated samples were characterized by scanning electron microscopy (SEM; Supra-55, Zeiss, Germany) with energy dispersive spectroscopy (EDS) and transmission electron microscopy (TEM, Talos F200x, Thermo Fisher Scientific, USA) equipped with EDS. In order to analyze the as-fabricated samples by TEM, several lamellae with dimensions of 5 μm × 8 μm × 0.1 μm were cut from the surface of the samples via the focused ion beam (FIB) (Scios dual beam, FEI, Portland, USA) technique.

## 3. Computational details

All density functional theory (DFT) calculations were implemented in the Vienna ab-initio simulation package (VASP) [38]. The projector-augmented wave (PAW) method with the exchange-correlation functional depicted by the general gradient approximation (GGA) from Perdew-Burke-Ernzerhof (PBE) was utilized to approximate electronic exchange and correlation effects [39,40]. The spin-polarization effects were included for 6-9$HEB_2$ because of the existence of the Cr element. The energy cutoff of the plane-wave basis was set to be 420 eV, and Γ-centered k-point meshes were used (9×9×4, 9×9×3, and 1×1×1 for 4$HEB_2$, 5$HEB_2$, and 6-9$HEB_2$, respectively) [41]. The energy convergence criterion of the electronic self-consistency cycle was $10^{-5}$ eV, and the atomic structure relaxed until the force on each atom was less than 0.01eV/Å. Supercells (108, 135, 162, 189, 192, and 108 atoms for 4-9$HEB_2$, respectively) were prepared based on the $ZrB_2$ conventional cell. The special quasi-random structure (SQS) approach, implemented in the Alloy Theoretic Automated Toolkit (ATAT) [42,43], was applied to generate the disordered occupations of the metal (Me) atoms in $HEB_2$. The integrated-crystal orbital Hamilton population (-ICOHP) was employed to analyze the strength of Me-Me, Me-B, and B-B bonds in $HEB_2$ [44].

Lattice distortion effect, one of the four core effects of high-entropy materials, has a significant influence on mechanical properties. Based on the quantitative analysis of lattice distortion in high-entropy alloy [45], the lattice distortion ($\bar{\mu}$) of $HEB_2$ can be derived as follows:

$$\bar{\mu} = \frac{2\sqrt{\sum_i^n (a_i^{eff} - \bar{a})^2} + \sqrt{\sum_i^n (c_i^{eff} - \bar{c})^2}}{3} \tag{1}$$

where $a_i^{eff}$ and $c_i^{eff}$ are the effective interatomic distances of the $i$-th element, $\bar{a}$ and $\bar{c}$ are the average interatomic distances of $n$ elements, and $n$ is the number of constituent metal elements in $HEB_2$. At the same time, the values of lattice distortion were also calculated from experiments by substituting the lattice constants for $\bar{a}$ and $\bar{c}$, which were acquired from XRD refinement and TEM. The effective interatomic distances of the $a_i^{eff}$ and $c_i^{eff}$ were calculated by:

$$a_i^{eff} = \sum_j^n f_j (1 + \frac{\Delta V_{ij}}{V_i})^{1/3} a_i \tag{2}$$

$$c_i^{eff} = \sum_j^n f_j (1 + \frac{\Delta V_{ij}}{V_i})^{1/3} c_i \tag{3}$$

where $f_j$ is the atomic fraction, $V_i$ is the atomic volume, $a_i$ and $c_i$ are the lattice constants of the individual diboride corresponding to the $i$-th element, and $\Delta V_{ij}$ is the change of the atomic volume of the $i$-th element dissolving into the $j$-th element. To calculate $\Delta V_{ij}$, $C_n^2$ kinds of binary diborides supercells were required to be built by SQS and $\Delta V_{ij}$ can be expressed as [46]:

$$\Delta V_{ij} = \frac{V_{before\ solid\ solution} - V_{after\ solid\ solution}}{N} \tag{4}$$

where the $V_{before\ solid\ resolution}$ is calculated by summing the atomic volume in the $i$-th binary diborides, $V_{after\ solid\ resolution}$ is the supercell volume after optimization, and the $N$ is the number of solvable elements corresponding to the half number of Me atoms in the diboride supercells. In addition, the electronegativity mismatch ($\delta Xr$) (Allred-Rochow scale) was estimated using the following equation [35]:

$$\delta Xr = \sqrt{\sum_i^n C_i(1 - \frac{Xr_i}{\bar{Xr}})^2} \quad (5)$$

where $C_i$ is the composition of the $i$-th metal element, $Xr_i$ is the electronegativity of the $i$-th element, and $\overline{Xr}$ is the average electronegativity.

## 4. Results and discussion

All $HEB_2$ samples are fabricated via the UHTS device, as exhibited in Fig. 1(a). The temperature vs. time profiles of the as-fabricated $HEB_2$ samples at different fabrication conditions are shown in Fig. 1(a). It can be clearly seen that the disk-shaped samples are fabricated by two different processing routes. The first processing route is attempted to fabricate all the 4-9$HEB_2$ samples, where the samples are initially heated to 1873 K at a heating rate of 50 K/s and held for 40 s, then the temperature of the system continues to rise to 2773 K at a heating rate of 65 K/s and keeps for 50 s to ensure the time for densification, and finally, the system temperature is gradually decreased to room temperature. The XRD patterns of the as-fabricated samples are depicted in Fig. 1(b). It is clear that only the main characteristic peaks of the single solid-solution diboride phase are detected in all samples without any diffraction peaks of impurities, indicating the successful synthesis of 4-9$HEB_2$ samples. The relative densities of the as-fabricated 4-6$HEB_2$ samples are very high, within the range of 98.0% ± 0.4% to 98.3% ± 0.5% (see Fig. 1(c)), while the relative densities of the as-fabricated 7-9$HEB_2$ samples are relatively low (~91-93%). To further improve the relative densities of 7-9$HEB_2$ samples, another processing route is explored by increasing the heating rates (55 K/s and 73 K/s) to get higher synthesizing temperatures (1973 K and 3073 K), as presented in Fig. 1(a). It is evident that single-phase 7-9HEB$_2$ samples are still successfully obtained without any impurity when the temperatures rise in both processes, as displayed in Fig. S1. The as-fabricated 7-9HEB$_2$ samples are high density, within the range of 97.9% ± 0.5% to 98.0% ± 0.5% (see Fig. 1(c).), suggesting that the as-fabricated 4-9HEB$_2$ samples are dense enough with similar relative densities. As a result, two different processing routes of UHTS methods are developed to fabricate the high-density 4-9HEB$_2$ samples within an ultrashort time (approximately $\leq$ 140 s). In addition, the Rietveld refined XRD patterns of the synthesized 4-9HEB$_2$ samples in Fig. S2 show that the as-fabricated samples are hexagonal structures ( space group *P6/mmc*) with low fitting parameters ($R_{wp}$ and $R_p$). Meanwhile, the refined unit cell parameters of the as-fabricated 4-9HEB$_2$ samples are in good accordance with the results from DFT calculations, as listed in Table 1, which further prove the successful synthesis of 4-9HEB$_2$ samples via the UHTS method. Fig. 1(d) displays the SEM image of the well-polished and acid-etched surface for the as-fabricated 9HEB$_2$ samples, from which the grain sizes are comparable, and the grain boundaries are apparent with a small number of pores. Moreover, the statistical average grain sizes of the as-fabricated 4-9HEB$_2$ samples after acid etching are fitted by the Gaussian function, as exhibited in Fig. 1(e). It is worth noting that the average grain sizes of the as-fabricated 4-9HEB$_2$ samples are in the range of 20.2 ± 1.8 to 27.5 ± 1.3 μm with an inconspicuous variation. Therefore, the influences of grain sizes and relative densities on the mechanical properties of the as-fabricated 4-9HEB$_2$ samples can be ignored.

As illustrated in Fig. S3, the homogeneity of metal elements of the as-fabricated 4-9HEB$_2$ samples was analyzed by the SEM images and the corresponding EDS compositional mappings. It can be seen that all metal elements are homogeneously distributed on the micron scale without any segregation or aggregation. Furthermore, each HEB$_2$ sample has almost identical atomic percentages of metal elements, well consistent with the stoichiometric ratio of the equal-atomic 4-9HEB$_2$ (Table S4), which further confirms the compositional homogeneity of the as-fabricated 4-9HEB$_2$ samples. The microstructure and elemental distributions of the as-fabricated 4-9HEB$_2$ samples at the nanoscale were further examined by TEM technology. Taking the as-fabricated 6HEB$_2$ and 9HEB$_2$ samples as examples, Figs. 2(a) and (b) present their high-resolution transmission electron microscopy (HRTEM) images and the corresponding fast Fourier transform (FFT) patterns inserted in the images, respectively. It can be seen that the grain boundaries for 6HEB$_2$ and 9HEB$_2$ samples are distinct and clear without any impurities or amorphous layers (marked with the yellow dashed line), indicating a favorable interface bonding. At the same time, it is obvious that the as-fabricated 6HEB$_2$ and 9HEB$_2$ samples have periodic lattice structures with two sets of fringes, where the interplanar spacings are measured to be 0.215 nm and 0.276 nm for 6HEB$_2$ and 0.210 nm and 0.274 nm for 9HEB$_2$, corresponding to {101} and {100} planes of metal diborides, respectively. The lattice parameters *a* and *c* are calculated to be 3.187 Å and 3.429 Å for 6HEB$_2$ and 3.164 Å and 3.269 Å for 9HEB$_2$, respectively, in good agreement with the results from XRD refinement (3.099 Å and 3.372 Å for 6HEB$_2$ and 3.079 Å and 3.310 Å for 9HEB$_2$, respectively) and DFT calculations (3.091 Å and 3.346

Å for $6HEB_2$ and 3.089 Å and 3.287 Å for $9HEB_2$, respectively), which indicates the successful synthesis of single-phase $6HEB_2$ and $9HEB_2$ samples. Furthermore, the scanning transmission electron microscopy (STEM) and the corresponding EDS mappings (Figs. 3(c)-(f)) of the as-fabricated $6HEB_2$ and $9HEB_2$ samples show that there are clean grain boundaries, and simultaneously, the distributions of all metal elements are highly homogeneous without any segregation or aggregation at the nanoscale. The atomic percentages of each metal element match well with the stoichiometric ratio of the corresponding $6HEB_2$ and $9HEB_2$ as listed in Table 2, which further verifies the successful synthesis of the single-phase $6HEB_2$ and $9HEB_2$. Based on these observations, it can be concluded that the as-fabricated 4-$9HEB_2$ samples are single-phase high-entropy solid solutions with uniform compositions and clean grain boundaries. Consequently, the impact of element distributions and grain boundaries on the mechanical properties of the as-fabricated 4-$9HEB_2$ samples can also be ignored.

The Vickers hardness of the as-fabricated 4-$9HEB_2$ samples as a function of applied load is shown in Fig. 3(a). It can be obviously observed that the Vickers hardness of each sample reduces gradually as the value of the applied load increases. These variations are attributed to the micropores left in the as-fabricated samples. Notably, the Vickers hardness exhibits an obviously increasing trend from $4HEB_2$ to $9HEB_2$ at the fixed load. That is to say, at an applied load of 1.0 kg, the as-fabricated $9HEB_2$ samples possess the highest Vickers hardness of 21.6 ± 0.8 GPa, while the as-fabricated $4HEB_2$ samples possess the lowest Vickers hardness of 16.5 ± 0.8 GPa. It should be noted that even though $CrB_2$, $MoB_2$, and $WB_2$ possess the lower hardness [47], the addition of these metallic elements (Cr in $6HEB_2$, Mo in $7HEB_2$, and W in $9HEB_2$) still induced an enhanced hardness, indicating that the chemical elements are not the influential factors on the changes of the hardness in $HEB_2$. In order to avoid the influence of tiny pores on the hardness, nanoindentation tests with more precise hardness were employed. As displayed in Figs. 3(b) and (c), it is obvious that the variations of the nanohardness and elastic modulus of the as-fabricated 4-$9HEB_2$ samples are well consistent with those in the Vickers hardness, where the value of the nanohardness is significantly enhanced with the increase of metal components. Additionally, the values of the nanohardness are higher than those of the Vickers hardness, of which the as-fabricated $9HEB_2$ samples possess the highest nanohardness (35.3 ± 0.6 GPa) and elastic modulus (617 ± 11 GPa) and the as-fabricated $4HEB_2$ samples has the lowest nanohardness (25.6 ± 0.6 GPa) and elastic modulus (530 ± 15 GPa). Combined with aforementioned morphological analyses, it can be deduced that changes in the measured hardness are not caused by elemental distributions, grain boundaries, grain sizes, or relative densities. Besides the hardness, the elastic modulus and the flexural strength of the as-fabricated 4-$9HEB_2$ samples were also investigated. As can be found in Fig. 3(d), the flexural strength is monotonically enhanced with the increase of incorporated elements. Particularly, the $9HEB_2$ samples possess the highest flexural strength (408 ± 11 MPa), approximately 20% higher than those of the $4HEB_2$ (343 ± 16 MPa). These observations confirm the hardening and strengthening effect of lattice distortion on the mechanical properties of $HEB_2$ by the addition of metal components.

In order to systematically explore the underlying hardening and strengthening mechanisms of $HEB_2$, some potential indicators are analyzed by DFT calculations, including lattice distortion, bond strengths, VEC, metallic states, and $\delta Xr$. Although lattice distortion has been postulated extensively in relation to the mechanical properties of HECs, the hardening and strengthening mechanisms of lattice distortion on the intrinsic mechanical properties have not been well investigated. Detailed constituent terms in Equation (1) are summarized in Table S5. As shown in Fig. 4(a), the values of lattice distortion of 4-9$HEB_2$ from DFT calculations are in good agreement with those from XRD and TEM calculations (The detailed TEM analysis of 4$HEB_2$ is displayed in Fig. S4), showing the accuracy of our DFT calculations. It is worth mentioning that there is a significant correlation between lattice distortion and hardness/flexural strength since they possess the same increasing tendency. To more intuitively describe the variance of lattice distortion in the 4-9$HEB_2$, the distribution of the 1-4 nearest neighbors (NN) interatomic distances for Me atoms and the 1-5 NN interatomic distances for B atoms of 4-9$HEB_2$ are calculated, and the results are shown in Figs. 4(b) and (c). It is evident that both the interatomic distances for Me and B atoms of 4-9$HEB_2$ distribute more widely with the increase of metal components, indicating the presence of a larger lattice distortion. In addition, the radius of metal elements and the calculated average in 4-9$HEB_2$ from the simple rule of mixtures are listed in Table S6. It is obvious that the tendency of average metal atomic radius changes in $HEB_2$ is inconsistent with the changes of lattice distortion. This result further verifies that the chemical elements have a minor impact on lattice distortion and hardness/strength. Hence, the enhancement of hardness and flexural strength in the $HEB_2$ mainly ascribes to the aggravation of lattice distortion.

To further verify the regulation of lattice distortion from the experiment (the atomic elastic strains in the lattices induced by lattice distortion), the geometrical phase analysis of different samples was applied. Here, we take the $4HEB_2$, $6HEB_2$, and $9HEB_2$ as comparable samples. Figs. 5 (a), (e), and (i) show the HRTEM images and corresponding fast Fourier transform (FFT) results of the $4HEB_2$, $6HEB_2$, and $9HEB_2$, respectively. Based on these results, their calculated atomic elastic strain distribution mappings (normal strain in [100] direction) are displayed in Figs. 5 (c), (g), and (k), respectively. It is notable that the strains of $HEB_2$ samples clearly show an increasing trend with the increase of metal components, implying the aggravation of lattice distortion. In general, a larger distortion can cause higher strains in the lattice, leading to a strain distribution farther away from zero strain and a wider full width at half maximum (FWHM) [48]. To better compare the strain values in the lattices of the $4HEB_2$, $6HEB_2$, and $9HEB_2$, their FWHM values were counted. As exhibited in Figs. 5 (d), (h), and (i), a continuously positive shift of strains with the increase of metal components can be observed. In other words, the FWHM values rise largely from 0.0316 to 0.0443 when the incorporated element number increases from 4 to 9. Therefore, the lattice distortion evaluated from the experiment is consistent with the results of the theoretical calculations, which proves the accuracy of the calculations on the lattice distortion.

The bond strength is one of the decisive factors that may significantly affect the hardness and flexural strength of materials. Associated with -ICOHP and charge densities, the impact of the different metal components on the bond strength of $HEB_2$ was further revealed. It is well known that B-B and Me-B bonds play a vital role in building hardness and strength in diborides [49]. As a result, the -ICOHP values of B-B and Me-B bonds in 4-9$HEB_2$ are calculated and displayed in Fig. 6(a). It can be found that the values of -ICOHP of the B-B bond are at least 3.5 times that of the Me-B bond, which suggests that the contribution of B-B bonds for building hardness and strength in $HEB_2$ is significantly larger than that of Me-B bonds. More importantly, with the increase of metal components, a significantly enhanced B-B bonding is observed in $HEB_2$, while the strength of Me-B bonds is almost unchanged. Meanwhile, the charge density distributions of the boron layer (002) in 4-9$HEB_2$ are shown in Fig. S5. As the number of components increases, the charge density between two B atoms increases apparently, implying the stronger B-B bonds, which further confirms the results of -ICOHP on B-B bonds. As a result, the promotion of hardness and flexural strength in $HEB_2$ is mainly beneficial from B-B bond strengthening. Detailed distributions of the -ICOHP of the B-B bonds within 4-9$HEB_2$ are shown in Fig. 6(b). It can be observed that the distributions of the -ICOHP of B-B bonds in 4-9$HEB_2$ are fluctuating, which indicates that the changes in B-B bond strength result from the variation of lattice distortion. In other words, the larger lattice distortion, the stronger B-B bond strength. Additionally, it can be found in Figs. 6(c)-(h) that the similarity in bond strengths of Me-B bonds in 4-9$HEB_2$ is related to the analogous -ICOHP distributions (0.6-1.8 eV), and the addition of Ti, Cr, and V atoms (possessing low –ICOHP values) is detrimental to the Me-B bond strength, whereas the addition of Hf, Mo, and W atoms has a positive impact, leading to an invariable Me-B bond strength in 4-9HEB$_2$. Therefore, the improvement of hardness and flexural strength in 4-9HEB$_2$ is achieved by B-B bond strengthening originating from lattice distortion.

VEC and $\delta Xr$ are another two important indicators that have been proposed to describe the hardness and flexural strength of materials [30,35,50]. Therefore, their influences are also taken into account. As shown in Fig. 7(a), it is clear that both VEC and $\delta Xr$ are not monotonically increased with the increase of metal components. Particularly, VEC drops dramatically in 5HEB$_2$ but is equivalent between 7HEB$_2$ and 8HEB$_2$. With respect to $\delta Xr$, it first rises gradually from 4HEB$_2$ to 7HEB$_2$ and then declines significantly from 7HEB$_2$ to 9HEB$_2$. As a consequence, VEC and $\delta Xr$ are not evidently correlated to the changes of hardness and flexural strength in 4-9HEB$_2$. In addition to lattice distortion, bond strength, VEC, and $\delta Xr$, metallic states are also considered important factors to affect the hardness and flexural strength of materials, where the fewer occupied electrons in metallic states, the higher the hardness and flexural strength of materials [36]. According to the crystal field theory [51], d orbits in HEB$_2$ can split into the $A_{1g}$ (d$z^2$) and two doubly degenerate $E_{1g}$ (dxz, dyz) and $E_{2g}$ (d$x^2$-$y^2$, dxy). The projected electron density of states (PDOS) of d orbits in 9HEB$_2$ is displayed in Fig. 7(b). It can be clearly found that the dxz and d$x^2$-$y^2$ overlap with dyz and dxy, respectively. At the same time, the overlap of d orbits in 4-8HEB$_2$ is consistent with that of 9HEB$_2$ (see Fig. S6). Therefore, the results of PDOS are in agreement with the crystal field theory, which is available for further study. The interactions of the p and splitting d orbits of B and Me atoms in $9HEB_2$ are exhibited in Fig. 7(c). The B-pz overlaps with $E_{1g}$ (dxz, dyz), suggesting the hybridization between them. Therefore, the remaining orbitals are metallic states since they do not participate in hybridization. The interactions of the p and splitting d orbits of B and Me atoms in $4\text{-}8HEB_2$ are shown in Fig. S7. It can be found that hybridization occurs between B-pz and $E_{1g}$ (dxz, dyz). As a result, the $A_{1g}$ ($dz^2$) and the $E_{2g}$ ($dx^2\text{-}y^2$, dxy) are metallic states of Me atoms in 4-9HEB2. By integrating below the Fermi level, the occupied electrons of the $A_{1g}$ ($dz^2$) and the $E_{2g}$ ($dx^2\text{-}y^2$, dxy) in $4\text{-}9HEB_2$ are obtained. As illustrated in Fig. 6(d), both the occupied electrons of the $A_{1g}$ ($dz^2$) and the $E_{2g}$ ($dx^2\text{-}y^2$, dxy) in $4\text{-}9HEB_2$ keep constant with the increase of metal components. Therefore, the changes of hardness and flexural strength in $4\text{-}9HEB_2$ are determined to be unimpacted by metallic states.

**5. Conclusion**

In summary, we have uncovered the hardening and strengthening mechanisms of HECs by investigating the mechanical properties of $4\text{-}9HEB_2$. To be specific, single-phase $4\text{-}9HEB_2$ samples with uniform compositions and clean grain boundaries without any impurities have been fabricated by UHTS method. Similar grain sizes and relative densities are obtained for the as-fabricated $4\text{-}9HEB_2$ samples. Both hardness and flexural strength have been measured to increase from $4HEB_2$ to $9HEB_2$. Based on DFT calculations, the lattice distortion of $HEB_2$ is found to impact hardness and flexural strength critically, where the escalation of lattice distortion helps to improve $HEB_2$'s hardness and flexural strength. Subsequent studies on the bond strengths of $4\text{-}9HEB_2$ show that the B-B bonds of $HEB_2$ are strengthened due to the aggravation of lattice distortion. Meanwhile, no evident correlations are observed in VEC, $\delta Xr$, and metallic states with hardness and flexural strength due to their irregular variations with the increase of metal components. Consequently, lattice distortion is identified to be the dominant indicator of hardness and flexural strength in $HEB_2$. Our results elucidate the hardening and strengthening mechanisms of HECs by intensifying lattice distortion.

**Acknowledgments**

We acknowledge the financial support from the National Key Research and Development Program of China (No. 2022YFB3708600), the National Natural Science Foundation of China (No. 52122204 and 51972116), and Guangzhou Basic and Applied Basic Research Foundation (No. 202201010632).

**Author Contributions**

Y. Chu conceived and designed this work. Y. Liu performed calculations. H. Tang and M. Ma performed experiments. Y. Chu, Y. Liu, H. Yu, and Z. Tang analyzed the data and wrote the manuscript. All authors commented on the manuscript.

**Notes**

The authors declare no competing financial interest.

**Table 1** Lattice parameters (*a* and *c*) of the as-fabricated 4-9$HEB_2$ samples from XRD, TEM, and DFT calculations.

| Samples | $a_{XRD}$ (Å) | $c_{XRD}$ (Å) | $a_{TEM}$ (Å) | $c_{TEM}$ (Å) | $a_{cal}$ (Å) | $c_{cal}$ (Å) |
|---|---|---|---|---|---|---|
| 4$HEB_2$ | 3.098 | 3.350 | 3.100 | 3.270 | 3.106 | 3.351 |
| 5$HEB_2$ | 3.108 | 3.362 | / | / | 3.112 | 3.382 |
| 6$HEB_2$ | 3.099 | 3.372 | 3.187 | 3.429 | 3.091 | 3.346 |
| 7$HEB_2$ | 3.092 | 3.355 | / | / | 3.106 | 3.313 |
| 8$HEB_2$ | 3.080 | 3.325 | / | / | 3.084 | 3.287 |
| 9$HEB_2$ | 3.079 | 3.310 | 3.164 | 3.269 | 3.089 | 3.287 |

**Table 2** Metal element atomic percentages of the as-fabricated $6HEB_2$ and $9HEB_2$ samples characterized by STEM-EDS.

| Samples | Elements (at.%) | | | | | | | | |
|---|---|---|---|---|---|---|---|---|---|
| | Ta | Nb | Ti | Zr | Hf | Cr | Mo | V | W |
| $6HEB_2$ | 18.36 | 15.89 | 16.24 | 17.62 | 16.45 | 15.44 | / | / | / |
| $9HEB_2$ | 13.25 | 10.84 | 11.14 | 10.42 | 9.38 | 12.39 | 11.25 | 10.80 | 10.53 |

## Figure captions

**Fig. 1.** Fabrication of 4-9$HEB_2$ samples via the UHTS method. (a) Temperature vs. time profiles at different fabrication conditions (The insets show a scheme of the UHTS device and the as-fabricated 9$HEB_2$ samples). (b) XRD patterns of the as-fabricated 4-9$HEB_2$ samples at the first processing route. (c) Relative densities of the as-fabricated 4-9$HEB_2$ samples. (d) SEM image of the well-polished and acid-etched surface for the as-fabricated 9$HEB_2$ samples. (e) Average grain sizes of the as-fabricated 4-9$HEB_2$ samples.

**Fig. 2.** TEM analysis of the as-fabricated 4-9$HEB_2$ samples. (a) and (b) HRTEM images of the as-fabricated 6$HEB_2$ and 9$HEB_2$ samples (The inset is the corresponding FFT patterns). (c-e) STEM images and the corresponding EDS compositional mappings of the as-fabricated 6$HEB_2$ and 9$HEB_2$ samples.

**Fig. 3.** Mechanical properties of the as-fabricated 4-9$HEB_2$ samples. (a) Vickers hardness as a function of the applied loads. (b) Nanohardness. (c) Elastic modulus. (d) 3-point flexural strength.

**Fig. 4.** Lattice distortion and interatomic distances of 4-9$HEB_2$ from DFT, XRD, and TEM. (a) Lattice distortion of 4-9$HEB_2$ from DFT, XRD, and TEM. (b) Distribution of the 1-4 NN interatomic distances for Me atoms of 4-9$HEB_2$. (c) Distribution of the 1-5 NN interatomic distances for B atoms in 4-9$HEB_2$. (Normalized numbers of atomic pairs are calculated from the number of certain atomic pairs divided by the maximum number of these atomic pairs.)

**Fig. 5.** TEM analysis of lattice distortion in the as-fabricated 4, 6, and $9HEB_2$ samples. (a-d) HRTEM images, corresponding fast Fourier transform (FFT), atomic elastic strain distribution mapping (normal strain in [100] direction), and FWHM distributions of $4HEB_2$. (e-h) HRTEM images, corresponding FFT, atomic elastic strain distribution mapping (normal strain in [100] direction), and FWHM distributions of $6HEB_2$. (i-f) HRTEM images, corresponding FFT, atomic elastic strain distribution mapping (normal strain in [100] direction), and FWHM distributions of $9HEB_2$. Different colors in the atomic elastic strain distribution mapping represent different strain values induced by the lattice distortion in a range of -0.2 to 0.2.

**Fig. 6.** Bond strength of B-B and Me-B bonds in 4-$9HEB_2$. (a) -ICOHP values of B-B and Me-B bonds in 4-$9HEB_2$. (b) Distribution of -ICOHP values of B-B bonds in 4-$9HEB_2$. (c-h) Distribution of -ICOHP values of Me-B bonds in 4-$9HEB_2$.

**Fig. 7.** Correlation between VEC, $\delta Xr$, and metallic states and components of 4-$9HEB_2$. (a) VEC and $\delta Xr$ values of 4-$9HEB_2$. (b) PDOS of d orbit in $9HEB_2$. (c) PDOS of Me and B atoms of $9HEB_2$. (d) Metallic states of 4-$9HEB_2$. The Fermi level is shifted to 0 eV.

# Figures

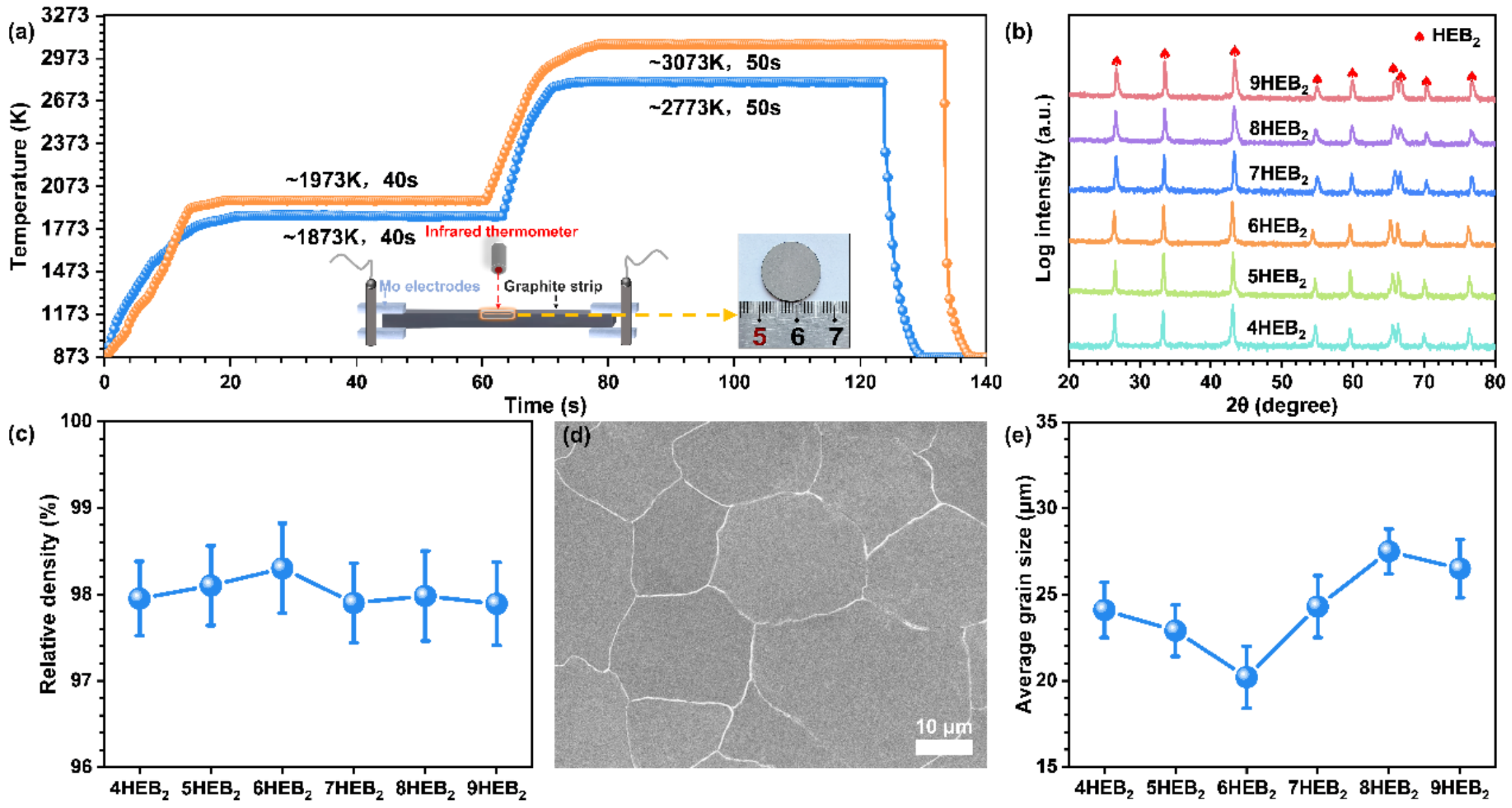

**Fig. 1.** Fabrication of 4-9HEB2 samples via the UHTS method. (a) Temperature vs. time profiles at different fabrication conditions (The insets show a scheme of the UHTS device and the as-fabricated 9HEB2 samples). (b) XRD patterns of the as-fabricated 4-9HEB2 samples at the first processing route. (c) Relative densities of the as-fabricated 4-9HEB2 samples. (d) SEM image of the well-polished and acid-etched surface for the as-fabricated 9HEB2 samples. (e) Average grain sizes of the as-fabricated 4-9HEB2 samples.

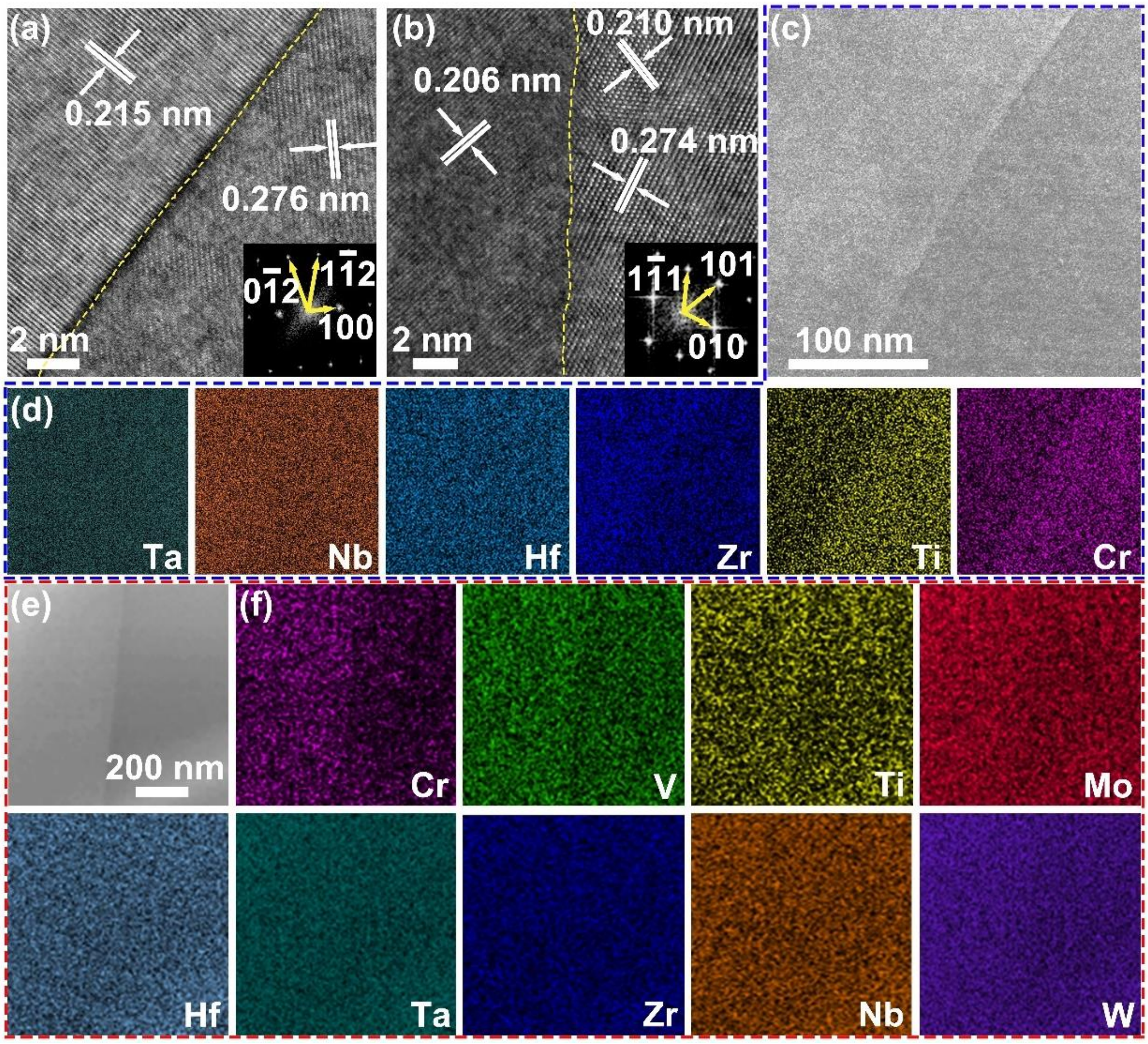

**Fig. 2.** TEM analysis of the as-fabricated 4-9HEB$_2$ samples. (a) and (b) HRTEM images of the as-fabricated 6HEB$_2$ and 9HEB$_2$ samples (The inset is the corresponding FFT patterns). (c-e) STEM images and the corresponding EDS compositional mappings of the as-fabricated 6HEB$_2$ and 9HEB$_2$ samples.

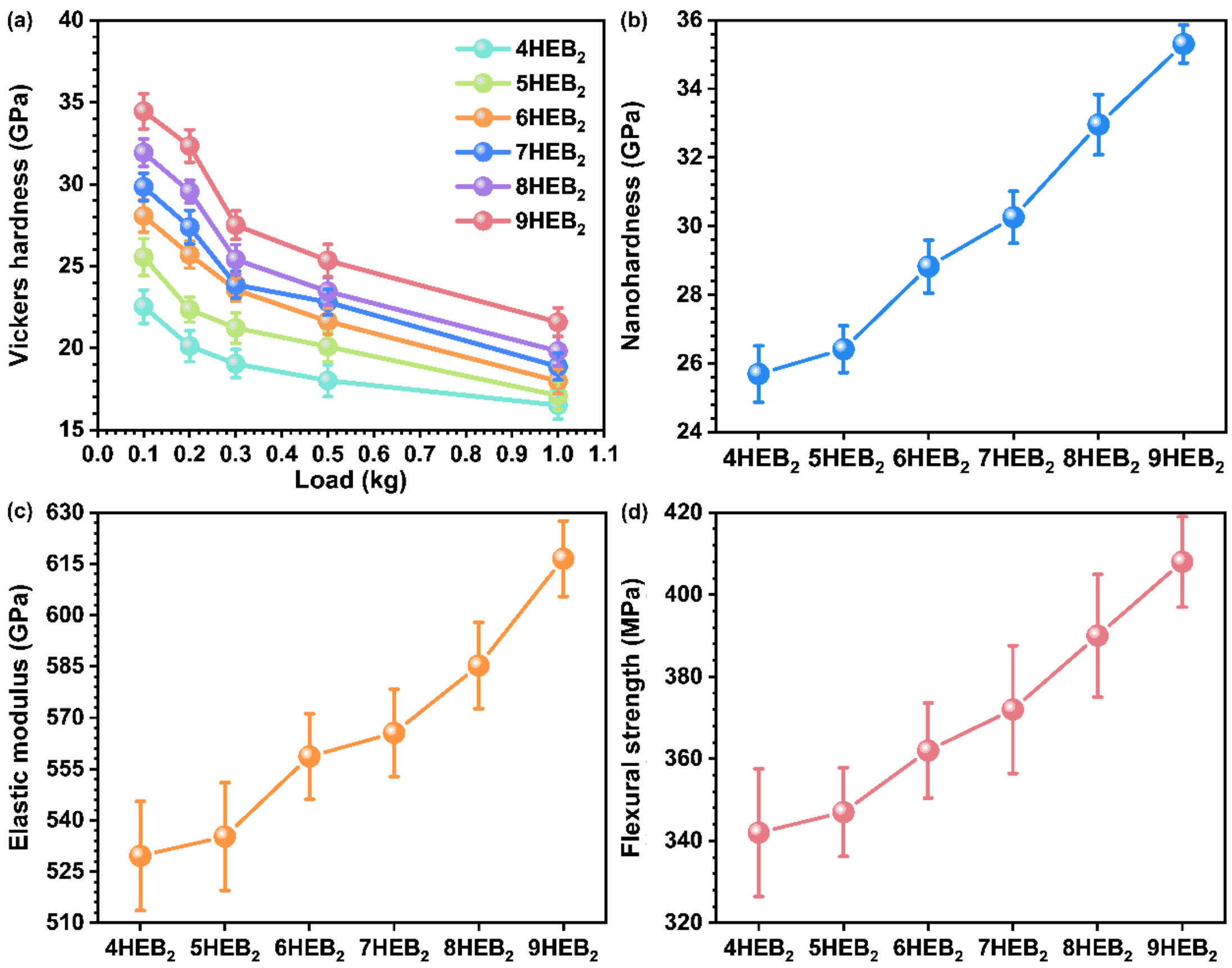

**Fig. 3.** Mechanical properties of the as-fabricated 4-9HEB$_2$ samples. (a) Vickers hardness as a function of the applied loads. (b) Nanohardness. (c) Elastic modulus. (d) 3-point flexural strength.

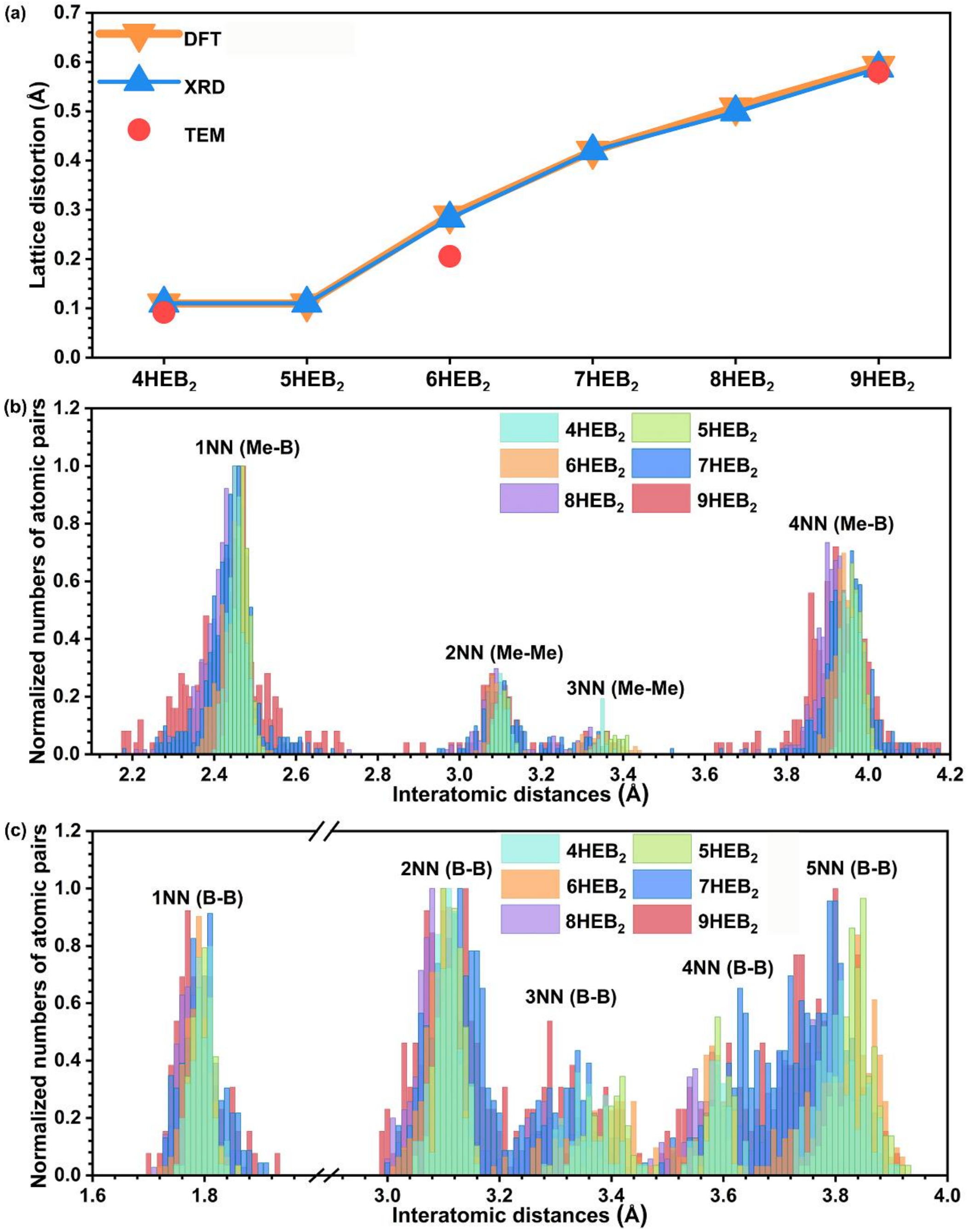

**Fig. 4.** Lattice distortion and interatomic distances of 4-9HEB$_2$ from DFT, XRD, and TEM. (a) Lattice distortion of 4-9HEB$_2$ from DFT, XRD, and TEM. (b) Distribution of the 1-4 NN interatomic distances for Me atoms of 4-9HEB$_2$. (c) Distribution of the 1-5 NN interatomic distances for B atoms in 4-9HEB$_2$. (Normalized numbers of atomic pairs are calculated from the number of certain atomic pairs divided by the maximum number of these atomic pairs.)

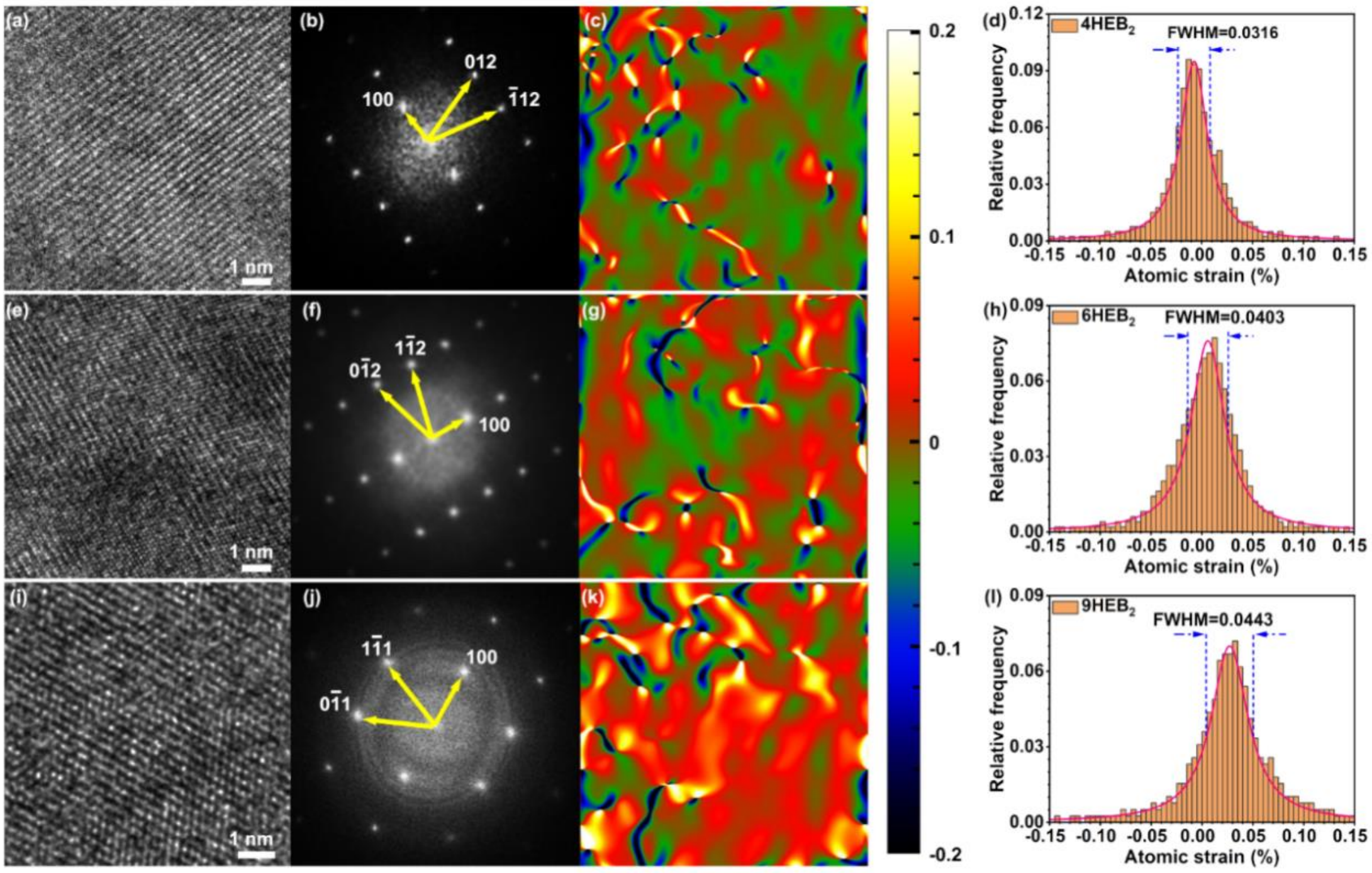

**Fig. 5**. TEM analysis of lattice distortion in the as-fabricated 4, 6, and $9HEB_2$ samples. (a-d) HRTEM images, corresponding fast Fourier transform (FFT), atomic elastic strain distribution mapping (normal strain in [100] direction), and FWHM distributions of $4HEB_2$. (e-h) HRTEM images, corresponding FFT, atomic elastic strain distribution mapping (normal strain in [100] direction), and FWHM distributions of $6HEB_2$. (i-f) HRTEM images, corresponding FFT, atomic elastic strain distribution mapping (normal strain in [100] direction), and FWHM distributions of $9HEB_2$. Different colors in the atomic elastic strain distribution mapping represent different strain values induced by the lattice distortion in a range of -0.2 to 0.2.

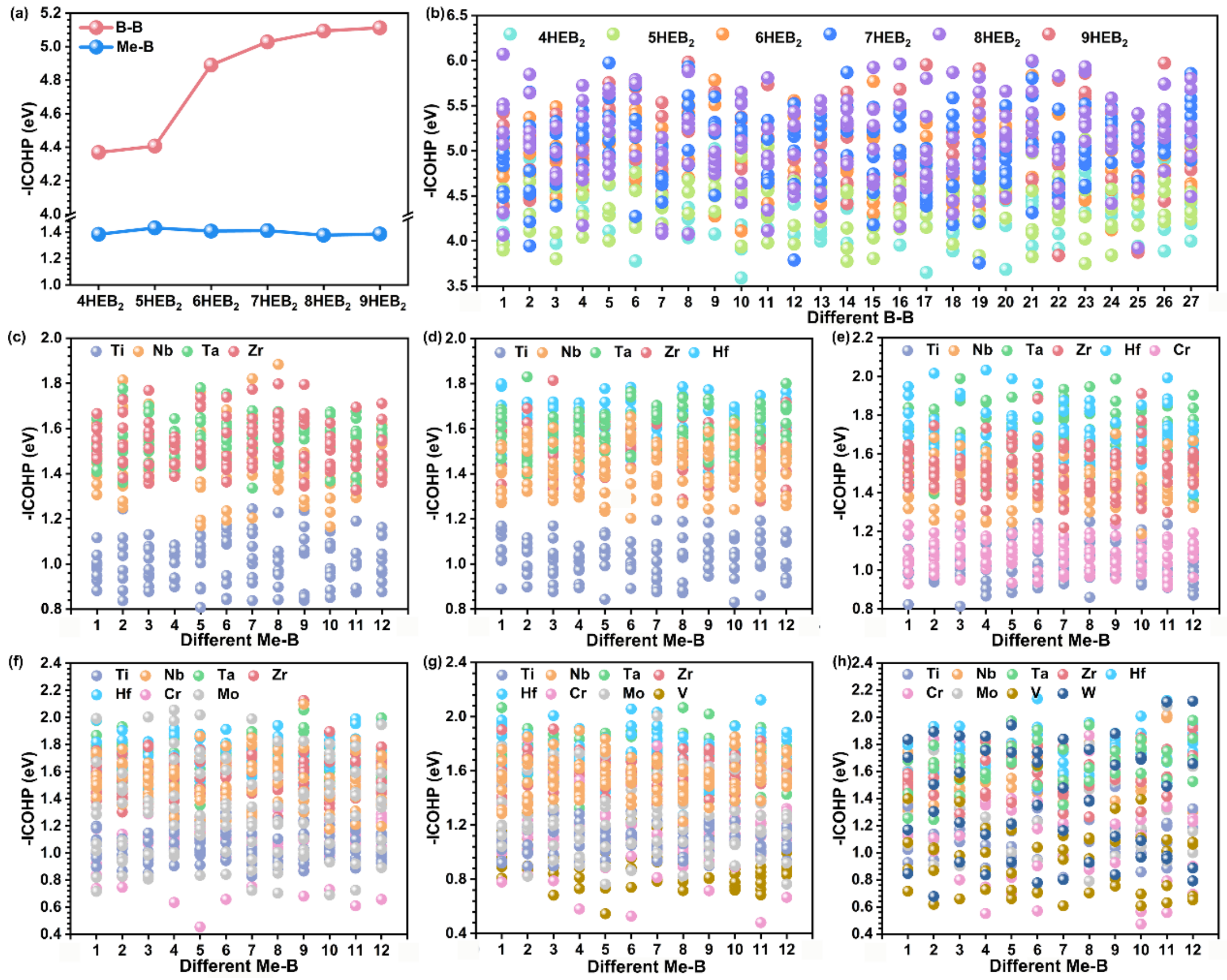

**Fig. 6.** Bond strength of B-B and Me-B bonds in 4-9HEB$_2$. (a) -ICOHP values of B-B and Me-B bonds in 4-9HEB$_2$. (b) Distribution of -ICOHP values of B-B bonds in 4-9HEB$_2$. (c-h) Distribution of -ICOHP values of Me-B bonds in 4-9HEB$_2$.

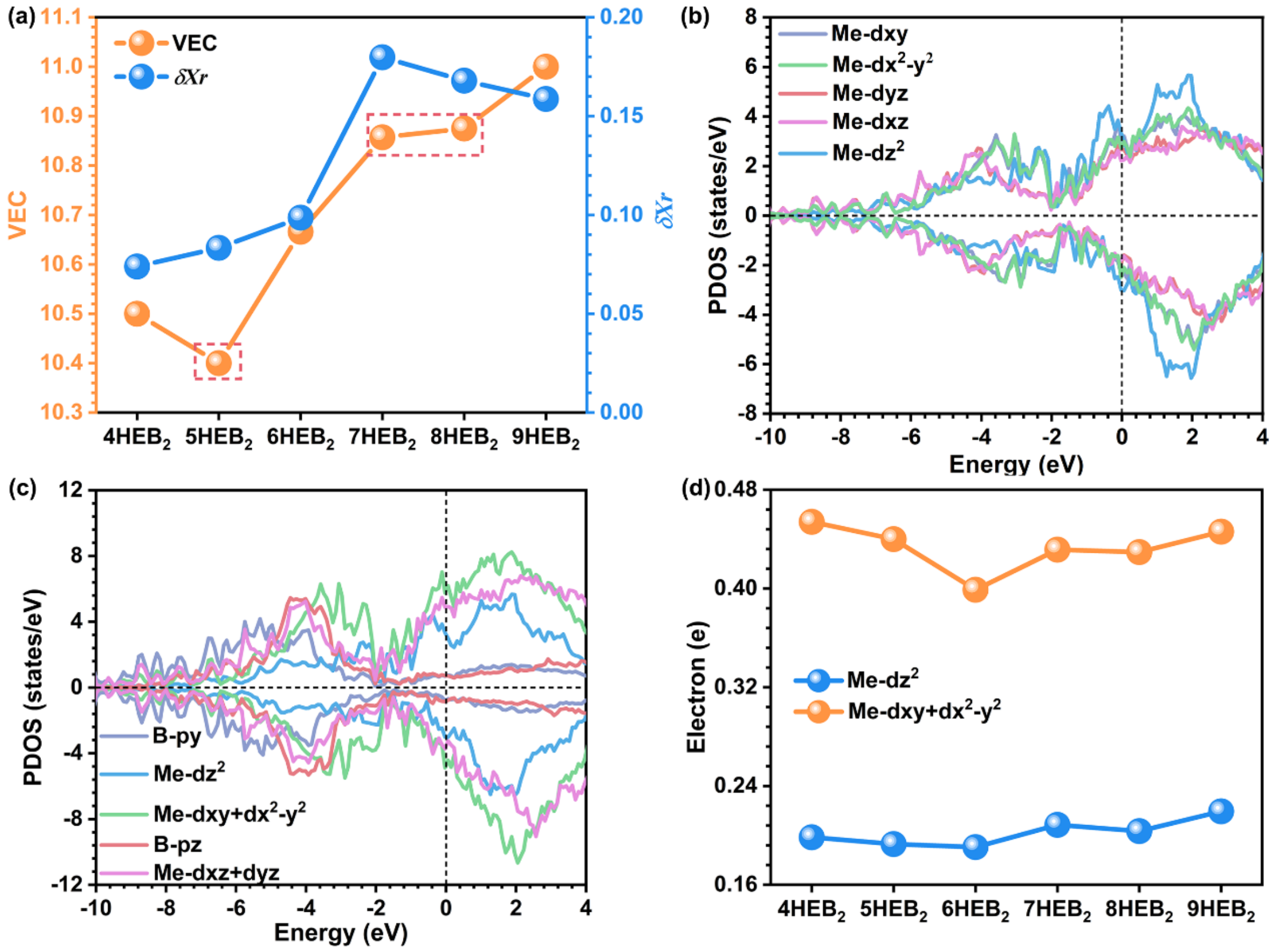

**Fig. 7.** Correlation between VEC, $\delta Xr$, and metallic states and components of 4-9HEB$_2$. (a) VEC and $\delta Xr$ values of 4-9HEB$_2$. (b) PDOS of d orbit in 9HEB$_2$. (c) PDOS of Me and B atoms of 9HEB$_2$. (d) Metallic states of 4-9HEB$_2$. The Fermi level is shifted to 0 eV.

**Supplemental Material for:**

# Unrevealing hardening and strengthening mechanisms in high-entropy ceramics from lattice distortion

Yiwen Liu[1], Haifeng Tang[1], Mengdong Ma[1], Hulei Yu, Zhongyu Tang, Yanhui Chu[*]

School of Materials Science and Engineering, South China University of Technology, Guangzhou, 510641, China

[1] These authors contribute equally to this article

[*] Corresponding author. chuyh@scut.edu.cn (Y. Chu)

**Table S1** Summary of the as-fabricated 4-9HE$B_2$ samples by UHTS.

| Samples | Components |
|---|---|
| 4HE$B_2$ | $(Ta_{1/4}Nb_{1/4}Ti_{1/4}Zr_{1/4})B_2$ |
| 5HE$B_2$ | $(Ta_{1/5}Nb_{1/5}Ti_{1/5}Zr_{1/5}Hf_{1/5})B_2$ |
| 6HE$B_2$ | $(Ta_{1/6}Nb_{1/6}Ti_{1/6}Zr_{1/6}Hf_{1/6}Cr_{1/6})B_2$ |
| 7HE$B_2$ | $(Ta_{1/7}Nb_{1/7}Ti_{1/7}Zr_{1/7}Hf_{1/7}Cr_{1/7}Mo_{1/7})B_2$ |
| 8HE$B_2$ | $(Ta_{1/8}Nb_{1/8}Ti_{1/8}Zr_{1/8}Hf_{1/8}Cr_{1/8}Mo_{1/8}V_{1/8})B_2$ |
| 9HE$B_2$ | $(Ta_{1/9}Nb_{1/9}Ti_{1/9}Zr_{1/9}Hf_{1/9}Cr_{1/9}Mo_{1/9}V_{1/9}W_{1/9})B_2$ |

**Table S2** Raw materials of the as-fabricated 4-9HEB$_2$ samples.

| Samples | Raw materials |
|---|---|
| 4HEB$_2$ | TaB$_2$, NbB$_2$, TiB$_2$, ZrB$_2$ |
| 5HEB$_2$ | TaB$_2$, NbB$_2$, TiB$_2$, ZrB$_2$, HfB$_2$ |
| 6HEB$_2$ | TaB$_2$, NbB$_2$, TiB$_2$, ZrB$_2$, HfB$_2$, CrB$_2$ |
| 7HEB$_2$ | TaB$_2$, NbB$_2$, TiB$_2$, ZrB$_2$, HfB$_2$, CrB$_2$, MoB$_2$ |
| 8HEB$_2$ | TaB$_2$, NbB$_2$, TiB$_2$, ZrB$_2$, HfB$_2$, CrB$_2$, MoB$_2$, VB$_2$ |
| 9HEB$_2$ | TaB$_2$, NbB$_2$, TiB$_2$, ZrB$_2$, HfB$_2$, CrB$_2$, MoB$_2$, VB$_2$, WB$_2$ |

**Table S3** Detailed information on the raw materials.

| Raw materials | Detailed information |
|---|---|
| $NbB_2$, $TiB_2$, $CrB_2$, $MoB_2$, $VB_2$, $WB_2$ | 99.5% purity, particle size: 1-3 μm, Qinhuangdao Yinuo High tech Materials Development Co., Ltd. Hebei, China |
| $TaB_2$, $ZrB_2$, $HfB_2$ | 99.9% purity, particle size: 1-3 μm, Shanghai ChaoWei Nanotechnology Co., Ltd. Shanghai, China |

**Table S4** Metal element atomic percentages of the as-fabricated 4-9$HEB_2$ samples characterized by SEM-EDS.

| Samples | Elements （at.%） | | | | | | | | |
|---|---|---|---|---|---|---|---|---|---|
| | Ta | Nb | Ti | Zr | Hf | Cr | Mo | V | W |
| 4$HEB_2$ | 23.52 | 23.85 | 26.28 | 26.35 | / | / | / | / | / |
| 5$HEB_2$ | 19.45 | 20.94 | 22.13 | 18.56 | 18.91 | / | / | / | / |
| 6$HEB_2$ | 16.82 | 17.05 | 17.21 | 18.14 | 16.38 | 14.40 | / | / | / |
| 7$HEB_2$ | 13.25 | 14.68 | 15.26 | 13.43 | 13.58 | 14.51 | 15.29 | / | / |
| 8$HEB_2$ | 11.89 | 12.21 | 13.54 | 13.27 | 12.31 | 13.48 | 11.45 | 11.85 | / |
| 9$HEB_2$ | 11.13 | 10.26 | 11.83 | 10.47 | 9.98 | 12.33 | 10.24 | 12.51 | 11.25 |

**Table S5** Effective lattice constants ($a_i^{eff}$ or $c_i^{eff}$) in each of constituent elements for 4-9$HEB_2$.

| | 4$HEB_2$ | 5$HEB_2$ | 6$HEB_2$ | 7$HEB_2$ | 8$HEB_2$ | 9$HEB_2$ |
|---|---|---|---|---|---|---|
| $a_{Zr}^{eff}$ | 2.97 | 3.16 | 3.30 | 3.40 | 3.47 | 3.53 |
| $c_{Zr}^{eff}$ | 3.33 | 3.54 | 3.69 | 3.81 | 3.88 | 3.95 |
| $a_{Ti}^{eff}$ | 2.94 | 3.12 | 3.25 | 3.35 | 3.41 | 3.47 |
| $c_{Ti}^{eff}$ | 3.13 | 3.33 | 3.46 | 3.57 | 3.64 | 3.70 |
| $a_{Ta}^{eff}$ | 3.06 | 3.26 | 3.41 | 3.52 | 3.59 | 3.66 |
| $c_{Ta}^{eff}$ | 3.28 | 3.26 | 3.65 | 3.77 | 3.85 | 3.93 |
| $a_{Nb}^{eff}$ | 3.09 | 3.28 | 3.43 | 3.55 | 3.62 | 3.69 |
| $c_{Nb}^{eff}$ | 3.32 | 3.30 | 3.70 | 3.82 | 3.90 | 3.97 |
| $a_{Hf}^{eff}$ | / | 3.11 | 3.25 | 3.35 | 3.42 | 3.48 |
| $c_{Hf}^{eff}$ | / | 3.46 | 3.61 | 3.72 | 3.80 | 3.87 |
| $a_{Cr}^{eff}$ | / | / | 3.53 | 3.65 | 3.72 | 3.79 |
| $c_{Cr}^{eff}$ | / | / | 3.61 | 3.73 | 3.80 | 3.88 |
| $a_{Mo}^{eff}$ | / | / | / | 3.58 | 3.65 | 3.73 |
| $c_{Mo}^{eff}$ | / | / | / | 3.96 | 4.04 | 4.12 |
| $a_{V}^{eff}$ | / | / | / | / | 3.61 | 3.39 |
| $c_{V}^{eff}$ | / | / | / | / | 3.64 | 3.71 |
| $a_{W}^{eff}$ | / | / | / | / | / | 3.48 |
| $c_{W}^{eff}$ | / | / | / | / | / | 3.17 |

**Table S6** Radius of metal elements and the calculated average in 4-9HEB$_2$ from the simple rule of mixtures.

| | Radius (pm) |
|---|---|
| Ta | 146 |
| Nb | 146 |
| Ti | 147 |
| Zr | 160 |
| Hf | 159 |
| Cr | 128 |
| Mo | 139 |
| V | 134 |
| W | 139 |
| 4HEB$_2$ | 150 |
| 5HEB$_2$ | 152 |
| 6HEB$_2$ | 148 |
| 7HEB$_2$ | 146 |
| 8HEB$_2$ | 145 |
| 9HEB$_2$ | 144 |

## Figures

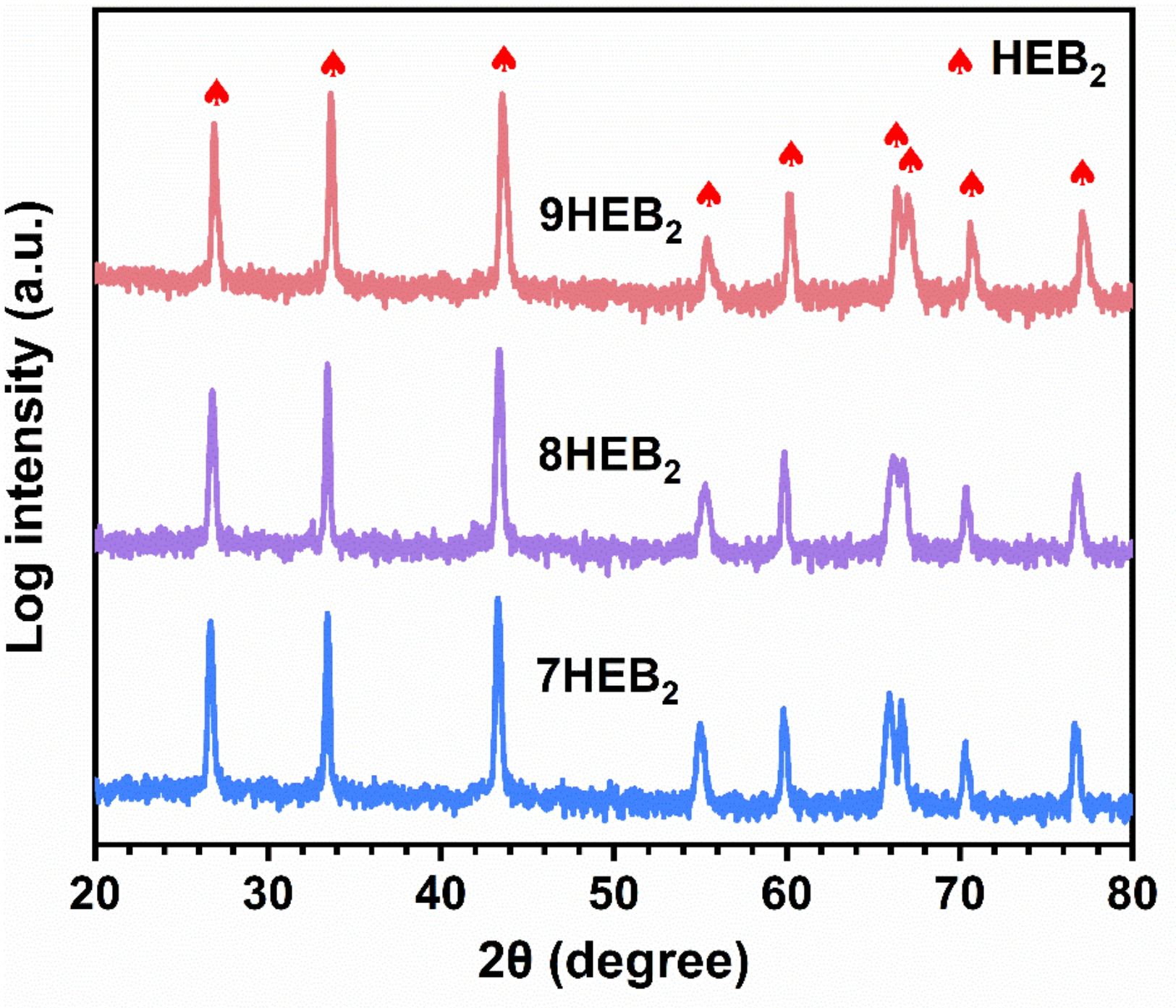

**Fig. S1.** XRD patterns of the as-fabricated 7-9HEB$_2$ samples at the second processing route.

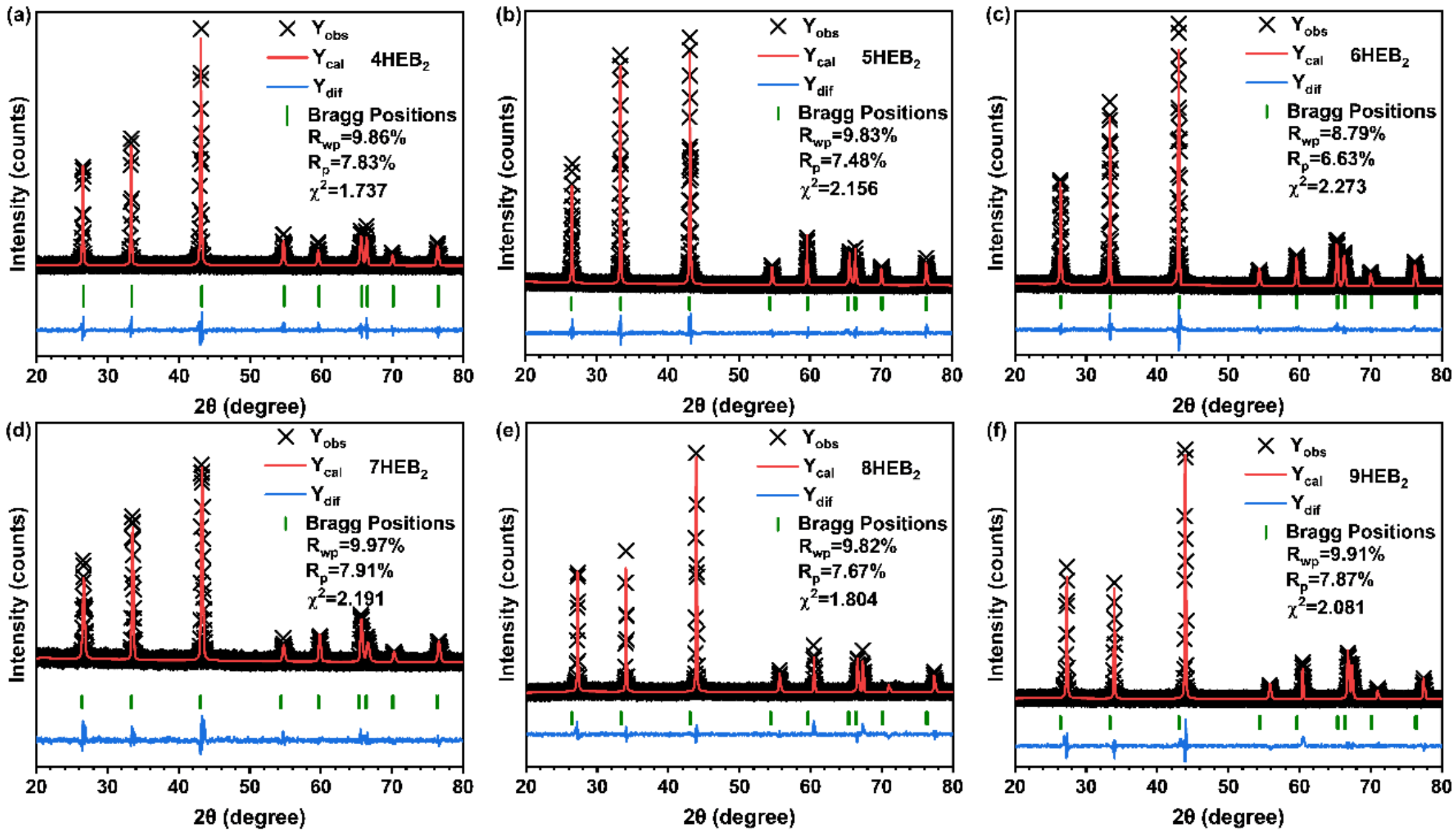

**Fig. S2.** XRD refinement patterns of the as-fabricated 4-9HEB$_2$ samples: (a) 4HEB$_2$, (b) 5HEB$_2$, (c) 6HEB$_2$, (d) 7HEB$_2$, (e) 8HEB$_2$ and (f) 9HEB$_2$.

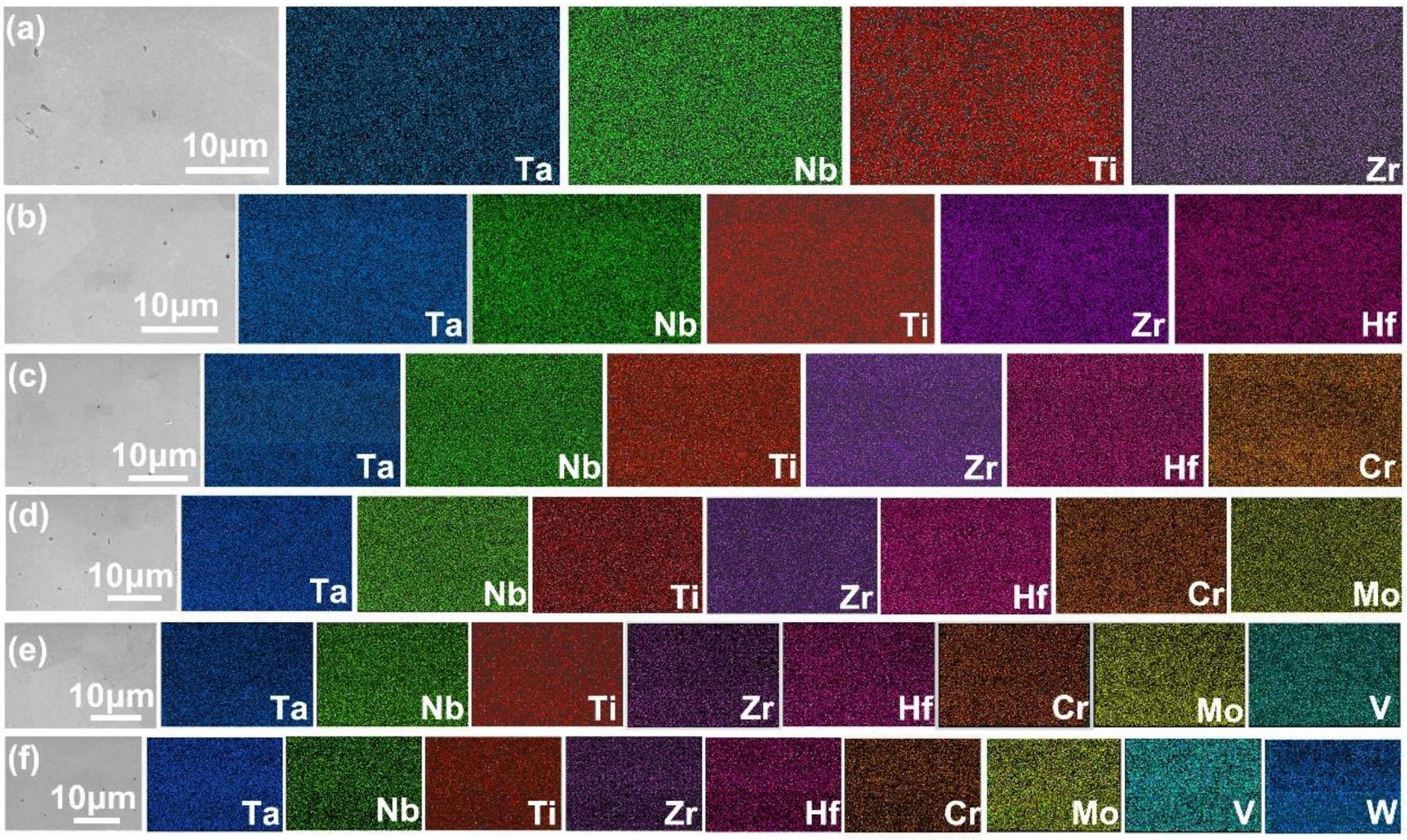

**Fig. S3.** SEM images and corresponding EDS compositional mappings of the as-fabricated 4-9HEB$_2$ samples.

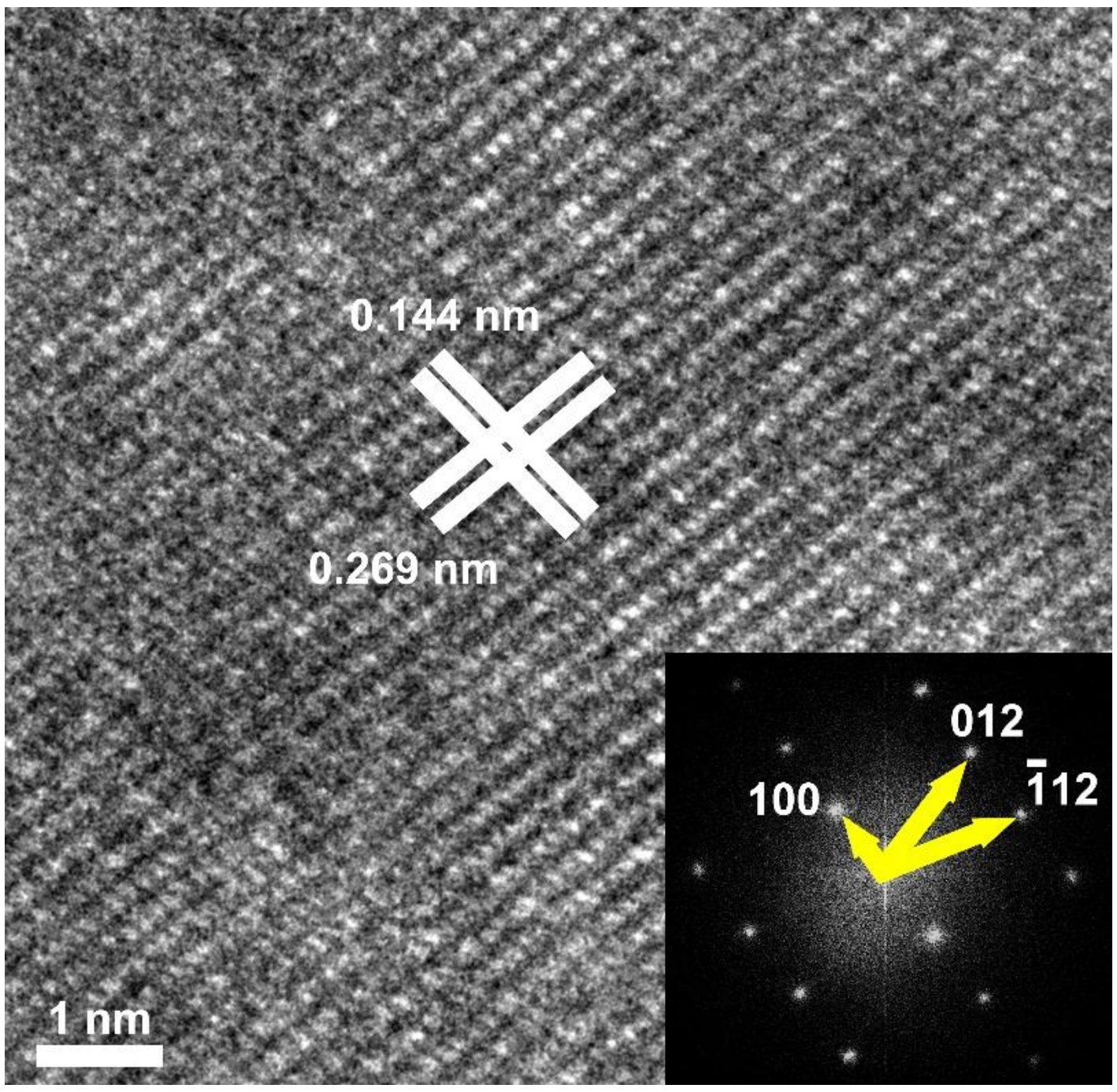

**Fig. S4.** HRTEM image of the as-fabricated $4HEB_2$ samples (The inset is the corresponding FFT pattern).

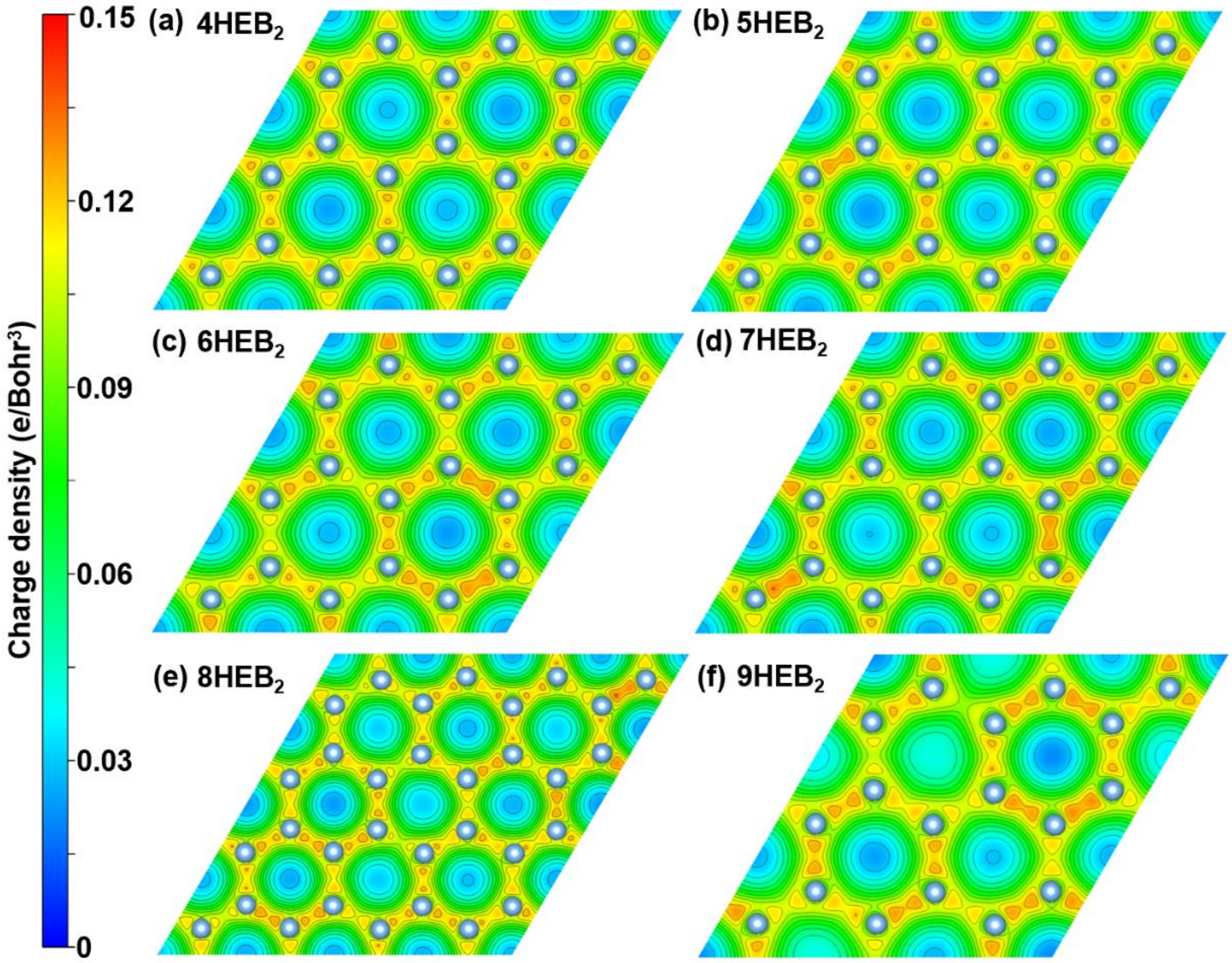

**Fig. S5.** Charge density distributions of the B layer (002) in 4-9HEB$_2$, where the blue atoms are B atoms.

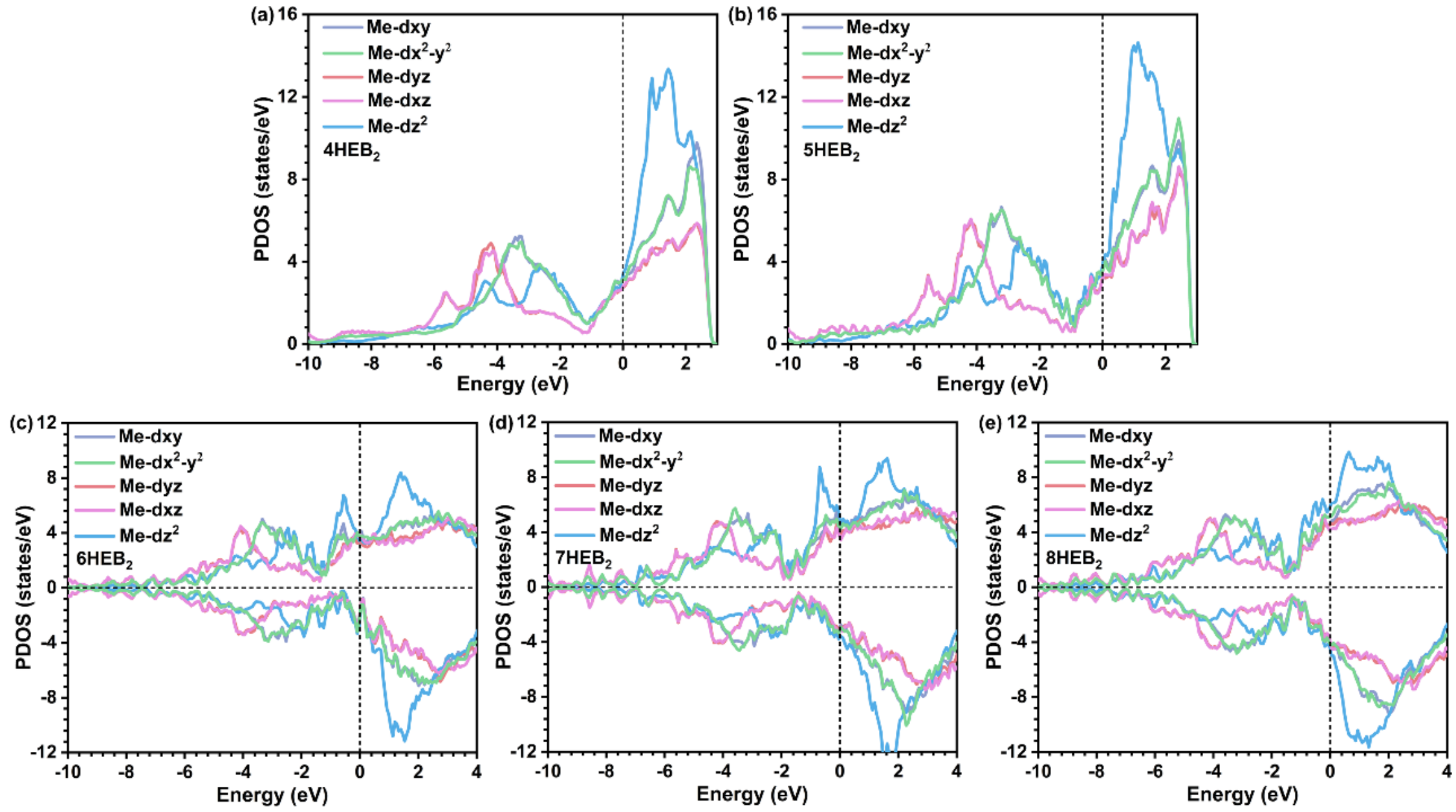

**Fig. S6.** PDOS of d orbits in 4-8$HEB_2$. The Fermi level is shifted to 0 eV.

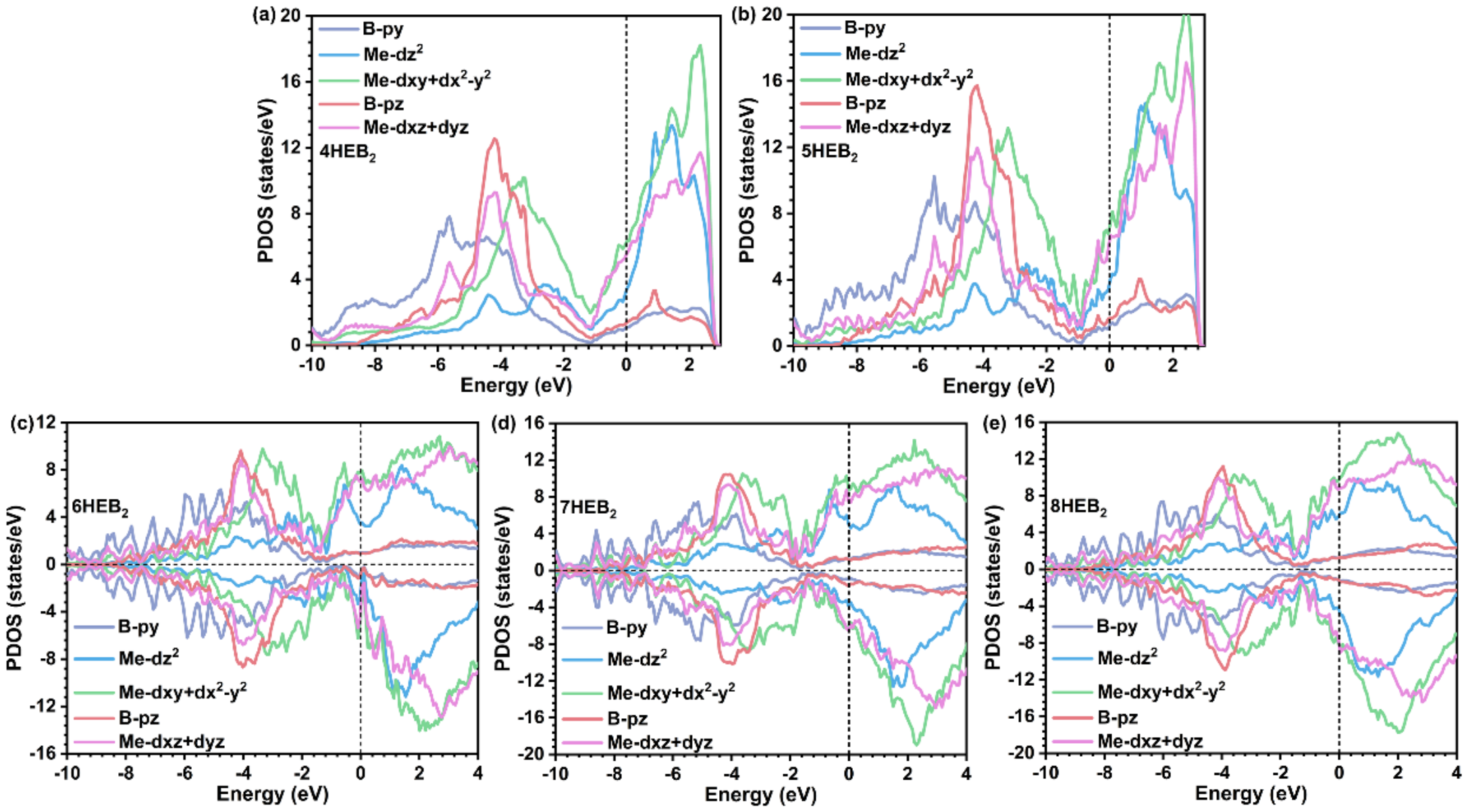

**Fig. S7.** PDOS of Me and B atoms of 4-8HEB$_2$. The Fermi level is shifted to 0 eV.